\documentclass[fleqn,twocolumn,10pt]{wlscirep}
\usepackage{xcolor}
\usepackage[utf8]{inputenc}
\usepackage[T1]{fontenc}

\usepackage{amsmath}
\usepackage{amssymb}
\usepackage{glossaries}
\usepackage{graphicx}
\usepackage{here}
\usepackage{times}

\usepackage{float}
\usepackage{booktabs}
\usepackage{bm}
\usepackage{siunitx}
\usepackage{adjustbox}

\usepackage{multirow}
\usepackage{mathtools}
\usepackage{xspace}

\usepackage{subcaption}
\usepackage[none]{hyphenat} 

\usepackage{tabularx}

\usepackage{hyperref}
\hypersetup{
    colorlinks=true,
    citecolor=orange,   
    urlcolor=blue,
    linkcolor=blue,
    filecolor=magenta,      
    urlcolor=cyan,
    breaklinks=false     
}

\newcommand{\vb}{$V_{\mathrm{B}}$}
\newcommand{\co}{CO}
\newcommand{\hh}{H$_{2}$}
\newcommand{\nn}{N$_{2}$}
\newcommand{\oo}{O$_{2}$}
\newcommand{\shh}{H$_{2}$S}
\newcommand{\coo}{CO$_{2}$}

\title{Nuclearity of Copper Clusters on hBN/SiC Heterostructure Modulates Molecular Adsorption}

\author[1]{Reza Khakpour~\thanks{Corresponding author: reza.khakpour@aalto.fi}}
\author[1,2,3]{Arsalan Hashemi}
\author[1]{Xiaoya Chang}
\author[4,5]{Nima Ghafari Cherati}
\author[2,3]{Mikko Karttunen}
\author[1,6]{Tapio Ala-Nissila~\thanks{Corresponding author: tapio.ala-nissila@aalto.fi}}

\affil[1]{MSP Group, Department of Applied Physics, Aalto University, P.O. Box 15600, FI-00076 Aalto, Espoo, Finland}
\affil[2]{European Laboratory for Learning and Intelligent Systems (ELLIS)  Institute Finland, Maarintie 8, 02150 Espoo, Finland}
\affil[3]{Department of Technical Physics, University of Eastern Finland, P.O. Box 1627, FI-70211 Kuopio, Finland}
\affil[4]{HUN-REN Wigner Research Centre for Physics, PO Box 49, H-1525 Budapest, Hungary}
\affil[5]{Department of Atomic Physics, Institute of Physics, Budapest University of Technology and Economics, M\H{u}egyetem rkp. 3, H-1111 Budapest, Hungary}
\affil[6]{Interdisciplinary Centre for Mathematical Modelling and Department of Mathematical Sciences, Loughborough University, Loughborough, Leicestershire LE11 3TU, United Kingdom}

\begin{abstract}
Defect engineering can transform inert two-dimensional (2D) materials into chemically active and electronically tunable platforms by creating anchoring sites for metal atoms and clusters.
Nevertheless, achieving precise control over the formation, thermodynamic and kinetic stability, electronic structure, and chemical reactivity of metal species confined at these defect sites remains a key unresolved challenge.
Here, we use density functional theory (DFT) calculations, assisted by machine-learning molecular dynamics (MLMD) simulations, to elucidate the stability, electronic structure, and reactivity of Cu clusters anchored at boron vacancy (\vb) sites in hBN/SiC heterostructures.
Systematic variation of the Cu-to-vacancy ratio reveals a transition from isolated Cu atoms to Cu$_{n>1}$@\vb\ clusters, with cluster growth directly reshaping stability, electronic structure, and surface reactivity.
Our results show that an individual \vb\ defect can be passivated by only three Cu atoms, which compensate the local charge deficiency and stabilize the defect through Cu--N coordination.
Capturing further Cu introduces localized midgap states that could control the reactivity of the Cu-decorated defect sites.
We probe the response of the Cu-decorated surface to chemically relevant gas adsorbates, \co, \hh, \oo, \nn, \shh, and \coo\ revealing their implications for surface reactivity and stability.
These gas-adsorption calculations reveal a pronounced cluster-size-dependent reactivity of Cu$_n$@\vb\ sites, where \co\ forms strong Cu--C bonds and \oo\ undergoes enhanced adsorption and molecular activation.
Overall, this work identifies defect-engineered hBN/SiC as a versatile 2D platform for stabilizing Cu clusters and tuning their gas-surface reactivity. By correlating Cu nuclearity at \vb\ sites with electronic structure, molecular activation, and environmental robustness, our findings provide design guidelines for nuclearity-dependent metal functionalization of 2D heterostructures.
These insights may serve as a basis for future developments in sensing, catalysis, and surface-chemistry correlation, while also supporting the design of durable single-atom like catalysts.

\end{abstract}

\begin{document}

\flushbottom
\maketitle


Decorating low-reactivity material surfaces with isolated or clustered transition metal (TM) atoms is an effective strategy to enhance surface reactivity while maximizing metal atom utilization.
Such metal centers can generate low-coordination, electronically tunable sites that are highly relevant to a wide range of applications, from catalysis to sensing~\cite{katiyar20232d,chen2023prediction,zheng2024growing}.
When combined with two-dimensional (2D) materials, the resulting hybrid systems are particularly attractive since they merge the structural flexibility of the substrate with the localized electronic and chemical characteristics of the metal species~\cite{thakur2024accelerating,liang2022progress,joudi2025two}.
Notable examples include nitrogen-doped graphene, which serves as a promising scaffold for anchoring TM single atoms and small clusters~\cite{liu2022dual,bord2023atomistic}.
However, these systems are often limited by the surface diffusion of TM species, which can destabilize isolated metal sites and promote undesired aggregation.
In addition to nitrogen-doped carbon materials, hexagonal boron nitride (hBN) has also been demonstrated to restrict TM mobility more effectively than graphene, thereby hindering nanocluster formation~\cite{Kohlrausch2025_AdvSci,Popov_nanolet_2023,Herrera-Reinoza2021_CM}.
Nevertheless, in currently explored systems, precise control over metal anchoring on the surface remains limited. Defects can migrate or the overall architecture lacks sufficient definition, complicating efforts to dictate metal placement, curb uncontrolled aggregation, and unambiguously identify well-defined active sites~\cite{zeng2026liquid,lang2019non,liu2024understanding}.

Similar to the existing paradigm of stabilizing single atoms by engineering supports that adjust the interactions between metal atoms and their solid-state substrates, we proposed a defective \emph{vertical} hBN/SiC heterostructure~\cite{Hashemi2026_PRM}.
Boron-vacancy (\vb) defects render the semiconducting hBN/SiC heterostructure an attractive 2D platform for the controlled anchoring of TM atoms.
The Moir\'{e} pattern appearing in hBN/SiC due to the lattice mismatch restricts the number of favorable \vb\ sites.
At these sites, the formation of interlayer chemical bonds reduces the structural energy, locally converting the van der Waals (vdW) interface into a chemically bonded structure. 
The under-coordinated nitrogen atoms around the vacancy site satisfy their dangling bonds by forming chemical bonds with the bottom layer SiC.
These locations can still function as potential trapping centers for metals that compensate for the local charge deficiency.
When only one or two metal atoms occupy these energetically favorable potential wells, they thereby passivate (i.e., neutralize the reactivity of) the \vb\ site. The resulting complex structures (Cu$_n$@\vb, where $n$ denotes the number of Cu atoms in the cluster) are anticipated to exhibit stronger binding and enhanced resistance to thermal fluctuations. This ultimately yields improved stability under subsequent reaction conditions ~\cite{Ghaderzadeh_acs_2026}.

Beyond the initial incorporation of metal atoms, their subsequent structural rearrangements and chemical transformations during synthesis and operation are equally important for the stability, activity, and functional response of metal-decorated surfaces.
Indeed, reactive atmospheric species can modify the local coordination environment, dispersion, oxidation state, and electronic structure of embedded metals, thereby transforming the activity of the targeted atomic sites.
This consideration is particularly critical for Cu-decorated surfaces, where oxygen-containing molecules may interact strongly with under-coordinated Cu species, triggering charge redistribution, oxidation, local structural rearrangement, and molecular activation~\cite{qi2023modulating,yang2022copper}.
Atomic-scale understanding of gas–surface interactions is therefore essential for distinguishing stable functionalization from surface degradation or uncontrolled activation.

For Cu-decorated hBN/SiC, resolving the stability–reactivity balance requires an atomistic understanding of how \vb\ defects capture and stabilize Cu species, thereby transforming the chemically inert hBN/SiC surface into tunable reactive sites.
Although \vb\ defects in hBN/SiC provide robust trapping centers for Cu atoms, the mechanisms modulating metal capture, stabilization, cluster nucleation, and the subsequent interaction of these sites with gas-phase species are yet to be explored.
This knowledge gap is particularly important since hBN/SiC is still an emerging heterostructure, despite growing experimental efforts toward the direct synthesis of hBN on Si- and SiC-based substrates~\cite{thomas2024colloidal,polley2023bottom,Ghaderzadeh_acs_2026,lin2022boron}.
Moreover, the recent experimental demonstration of monolayer SiC further underscores the feasibility and relevance of investigating well-defined hBN/SiC interfaces~\cite{lin2012light,polley2023bottom,thomas2024colloidal}.
A systematic understanding of these systems is therefore essential. However, this is experimentally challenging because vacancy density, metal loading, defect distribution, cluster formation, and adsorbate interactions create a large configurational space that is difficult to isolate and quantify by experiment alone.

Here, we extend our previous work in Ref. \cite{Hashemi2026_PRM} on transition-metal stabilized hBN/SiC to the case of Cu adatoms and clusters. We combine density functional theory (DFT) with machine-learning molecular dynamics (MLMD) to reveal how \vb\ defects in hBN/SiC stabilize Cu atoms and clusters, forming Cu$_n$@\vb\ active sites with size-dependent electronic structures and chemical reactivity toward gas adsorption.
We first use MLMD simulations to map Cu capture and clustering as a function of the Cu/\vb\ ratio.
We then analyze representative Cu$_n$@\vb\ motifs to determine how Cu nuclearity modifies defect-localized states and structural stability.
Finally, adsorption calculations using selected gas molecules demonstrate how both the cluster size and the metal coordination environment determine adsorption strength, charge transfer, and the activation of molecular bonds.

\section*{Results and discussion}
\label{sect:results_discussion}

\subsection*{From single-atom traps to copper clusters}
\label{subsec:cuadsorptionMD}

%
\begin{figure*}[ht!]
    \centering
    \includegraphics[width=0.95\textwidth]{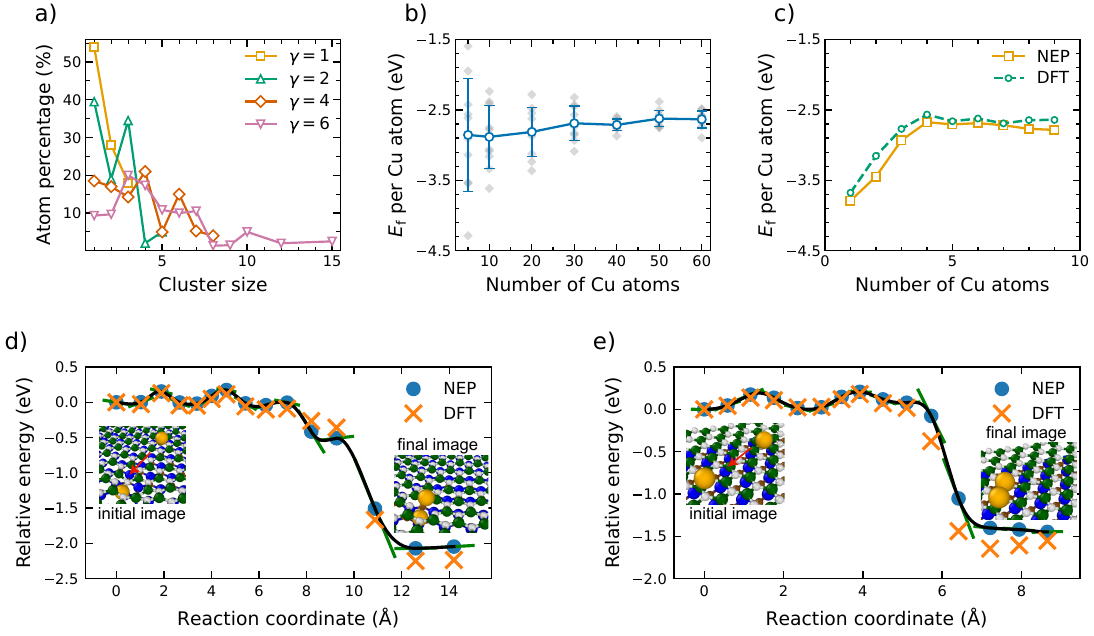}
    \caption{(a) Cu cluster-size distributions averaged over ten independent simulations, at $T = 300 \rm{~K}$, for different Cu-to-vacancy ratios $\gamma = N_{\rm{atom}}/10$. (b) Formation energy per Cu atom as a function of Cu loading for systems containing 10 \vb. Individual configurations are shown as faint symbols, and mean values are shown with standard-deviation error bars. (c) NEP and DFT comparison of the formation energy per Cu atom for Cu decoration of a single \vb. NEB-calculated reaction profiles for the binding of two Cu atoms on (d) \vb-containing and (e) pristine hBN/SiC. DFT results were obtained using the PBE xc-functional.}
    \label{fig:Cluster_num_ave}
\end{figure*}
Before becoming trapped at vacancy sites, Cu atoms undergo surface diffusion across the hBN substrate, exhibiting a preferential occupation of adsorption sites located above N atoms in pristine regions.
Once trapped, Cu becomes confined within the \vb\ defect and remains nearly immobile.
The atomic trajectories over the final 100~ps (200 frames) were examined to evaluate whether the Cu atom became confined.
To this end, the root mean squared displacement (RMSD) of the $x$-coordinate for a single Cu atom is given by
\begin{equation} 
    \mathrm{rmsd}(x) = \sqrt{\frac{1}{N_f}\sum_{i=1}^{N_f} (x_i - \bar{x})^2},
\end{equation}
where $N_f$ is the total number of frames used in the calculation, $x_i$ is the $x$-coordinate at the $i_{\rm{th}}$ frame, and $\bar{x} = (\sum_{i=1}^{N_f} x_i)/N_f$ is the mean $x$-coordinate. 
Similarly, the $\mathrm{rmsd}(y)$ and  $\mathrm{rmsd}(z)$ can be determined, and each component would define the corresponding tracer diffusion coefficient $D^\alpha_{\rm T} = \lim_{t \to \infty} {\rm rmsd(\alpha})^2/6t$ in the hydrodynamic limit with no trapping, with $\alpha=x,y,z$. 
The average is defined by $(\mathrm{rmsd}(x) + \mathrm{rmsd}(y) + \mathrm{rmsd}(z))/3$. 

Our simulations across different Cu/$V_{\mathrm{B}}$ ratios consistently indicate that there is indeed Cu trapping which can be quantified by the behavior of the RMSD, which describes localized vibrations instead of diffusion.
The results for ten independent simulations are shown in Fig.~S1 of Supplementary Information (SI).
In each case, ten \vb\ defects were randomly introduced into the hBN layer.
The maximum atomic vibration amplitudes remain below 3~\AA, indicating that Cu atoms are effectively immobilized on the hBN surface by the end of the trajectories and that a 5~ns simulation time is sufficient to reach trapped configurations.
With increasing Cu/\vb\ ratio, the vibration amplitudes increase slightly, which can be attributed to the formation of larger vacancy-anchored Cu clusters and, at higher Cu loadings, the presence of more weakly confined Cu atoms.

Next, we identified the cluster size using an interatomic distance criterion implemented in the MDAPY~\cite{wu2023mdapy} package.
More specifically, atoms $i$ and $j$ are considered to belong to the same cluster if $\lVert \mathbf{r}_i - \mathbf{r}_j \rVert<r_{\text{cut}}$, where $\mathbf{r}_i = (x_i, y_i, z_i)$ denotes the coordinate of the $i_{\rm{th}}$ atom, $\lVert \mathbf{r}_i - \mathbf{r}_j \rVert$ is the Euclidean distance between atom $i$ and $j$.
The cutoff distance was set to $r_{\text{cut}} = 3$~\AA, which is larger than the first-neighbor distance in the radial distribution function of bulk Cu (2.55~\AA~\cite{lee2003semiempirical}).

Our calculations indicate that Cu atoms preferentially aggregate into larger clusters, especially as the Cu concentration increases.
The generated Cu cluster sizes and their corresponding atom percentages at room-temperature ($T = 300~\rm{K}$) are summarized in Fig.~\ref{fig:Cluster_num_ave}a. 
Due to negligible variations across ten independent simulations, the averaged results are presented to reduce the influence of initial configuration differences.
Here, the atom percentage of each cluster size is calculated as the product of the cluster size ($n$) and its average number ($\bar{N}_n$), divided by the total number of Cu atoms ($N_\mathrm{atom}$), $i.e.$, $n \bar{N}_n \times 100$ / $N_\mathrm{atom}$.

In the low Cu/$V_{\mathrm{B}}$ regime, however, the captured Cu atoms predominantly appear as isolated atoms or as small clusters containing at most three atoms.
As the Cu/$V_{\mathrm{B}}$ ratio increases, the fraction of isolated Cu atoms gradually decreases, while the tendency to form larger clusters correspondingly increases.

The energetic favorability of Cu nucleation is evaluated using the defect-formation energy for different Cu densities on hBN/SiC containing ten \vb\ defects (Fig.~\ref{fig:Cluster_num_ave}b).
This process is exothermic, and the growth of particles is controlled by the drive to minimize the overall surface energy.
At low Cu loading, for example $N_\mathrm{atom} = 5$, the formation energies show substantial dispersion.
This variation arises from differences in the distribution of Cu atoms, with configurations in which isolated Cu atoms are captured by \vb\ sites being energetically more favorable than those where Cu atoms remain on pristine surface regions. As Cu loading increases, more dangling bonds are passivated, resulting in more consistently favorable configurations.

To estimate the Cu loading necessary to passivate an individual defect site, we evaluated the binding strength by calculating the formation energies of Cu clusters with varying $N_\mathrm{atom}$ on a single \vb\ defect (Fig.~\ref{fig:Cluster_num_ave}c).
The results suggest that three Cu atoms are sufficient to passivate the \vb\ site.
This will be examined in greater detail in the subsequent section.
Beyond this cluster size, the binding energy tends to saturate approximately at $-2.6$~eV per Cu, indicating a transition from defect-mediated stabilization to Cu–Cu metallic interactions associated with the growth of larger Cu clusters. 
We further benchmarked the NEP model against DFT calculations and found close agreement, supporting the reliability of the large-scale simulations based on this potential.

To further elucidate the role of \vb\ in trapping Cu atoms and also mobility of Cu on the hBN surface, we used nudged elastic band (NEB) method~\cite{MILLS1995305} (Figs.~\ref{fig:Cluster_num_ave}d and e).
Cu migration proceeds across B--N bonds with a low activation barrier of approximately 200~meV, while the energy minima correspond to configurations in which Cu occupies sites above N atoms.
Cu--Cu aggregation is effectively barrierless on both pristine and defective hBN.
Notably, when a Cu atom is already trapped at \vb, the aggregation reaction energy is approximately 0.55~eV more favorable than on the pristine surface. This enhanced thermodynamic driving force indicates that the charge deficiency associated with \vb\ plays a critical role in capturing additional Cu atoms and promoting cluster nucleation.
Finally, the close agreement between the DFT and NEP energies across the entire migration pathways demonstrates the robustness and transferability of the trained interatomic potential.

\subsection*{Electronic structure properties}
\label{subsec:cuadsorption}

\begin{figure}[htbp!]
    \centering
    \includegraphics[scale=0.65]{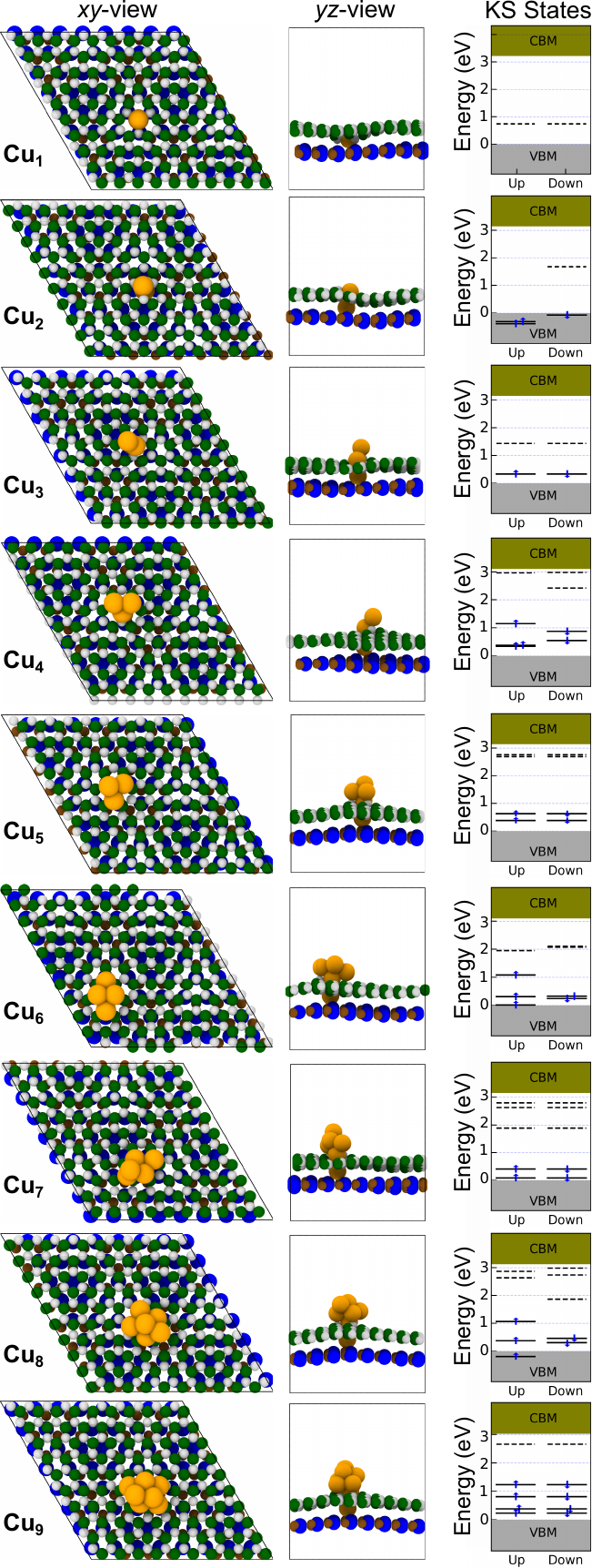}
    \captionsetup{justification=justified, singlelinecheck=false}
    \caption{Top ($xy$) and side ($yz$) views of Cu clusters on hBN/SiC heterostructure, along with the corresponding ground-state Kohn-Sham (KS) energy levels. Cu, Si, C, B, and N atoms are shown in orange, blue, brown, green, and light gray, respectively. Occupied KS states are shown as solid lines with $\uparrow$ and $\downarrow$ indicating spin-up and spin-down, respectively, while unoccupied states are represented by dashed lines. Energy levels at or below the valence band maximum (VBM) are shown in gray, while those at or above the conduction band minimum (CBM) are shown in olive. DFT results were obtained using the HSE03 xc-functional.}
    \label{fig:fig-str-orb}
\end{figure}

Figure~\ref{fig:fig-str-orb} summarizes the relaxed structural configurations and ground-state electronic structures of Cu$_n$@\vb\ complexes, where $1 \leq n \leq 9$.
The pristine \vb\ defect exhibits a doublet ground state, as discussed in Ref.~\citeonline{Hashemi2026_PRM}.
Upon Cu adsorption, charge donation from the Cu atoms shifts the defect-induced electronic levels and modifies the spin state of the system.
All the complexes relax to the lowest-symmetry configuration, namely $C_1$.
Notably, multiple Cu atoms gather to form a mushroom-like atomic cluster, where one atom resides in the interlayer vdW gap while the remaining atoms are located above \vb.
%

For low Cu coverages ($n = 1-3$), one Cu atom migrates into the vdW gap, where it predominantly bonds to the three nearest N atoms in the hBN layer and forms a bond with a neighboring C atom in the SiC monolayer.
For Cu$_1$, the Cu atom donates its $s$-orbital electron to a deep substrate defect state that becomes resonant with the valence band, resulting in a singlet closed-shell system.
The remaining empty in-gap state has antibonding Cu--N $\pi$-character, with additional contributions from Cu--C antibonding interactions.
Metal adsorption weakens the interlayer Si--N interaction and slightly elongates nearby Si--C bonds.
Upon adding a second Cu atom, the interlayer Si--N bond is broken, and Cu--Cu interactions emerge, associated with a reorganization of Cu--N coordination.
This system adopts a doublet spin state, with the occupied states lying within the valence band.
With the addition of the third Cu atom, one defect level within the band gap remains occupied while the other remains empty. 
The occupied state exhibits clear Cu--N bonding character with significant Cu $d$-orbital contributions, whereas the empty state is antibonding and localized on the most weakly coordinated Cu atom.
The binding energies per Cu atom for clusters containing one, two, and three Cu atoms are $-3.0$, $-3.07$, and $-2.62$~eV, respectively.
The disappearance of the original vacancy state at Cu$_2$ and the reappearance of gap states at higher Cu coverage reflect two different regimes, namely Cu-induced passivation at low coverage and Cu-cluster-induced gap-state formation at higher coverage.

For larger clusters ($n \ge 4$), the electronic trends distinctly vary for even- and odd-$n$ complexes. 
The even-$n$ systems remain in an open-shell doublet state.
Upon increasing the cluster size by two Cu atoms, two previously occupied defect-derived states are pushed into the valence band, while two new states emerge within the band gap. 
As a result, Cu$_4$, Cu$_6$, and Cu$_8$ show qualitatively similar KS states.
In contrast, the odd-$n$ complexes favor closed-shell singlet states. 
In these systems, the additional Cu-derived electron does not remain as a localized unpaired spin, but is instead paired through Cu--N and Cu--Cu hybridization. 
At the large clusters, stronger Cu--Cu coupling leads to the formation of a more cluster-like electronic manifold. 
The splitting of this manifold produces several Cu-derived bonding and weakly antibonding states, more than one of which lies inside the band gap.
Therefore, the additional occupied gap states in Cu$_9$ signal the onset of Cu-cluster-induced electronic states, rather than a simple continuation of the single-vacancy-state picture.
Overall, the cluster-substrate interaction weakens.
We, indeed, attribute this to the saturation of the defect-induced stabilization by the first three Cu atoms.
Upon further Cu incorporation, the additional atoms interact more strongly with the existing Cu cluster than with the defective substrate, leading to a more isolated cluster-like configuration.

The emergence of Cu-induced midgap states enables electronic excitations in the near-infrared to visible spectral range across all Cu-decorated systems.
This indicates that Cu functionalization introduces an additional optical functionality to the otherwise semiconducting hBN/SiC heterostructure.

Bond analysis indicates that the B--N bonds in hBN (1.41--1.48~\AA) represent the strongest covalent interactions.
In the SiC substrate, Si--C bond lengths range from 1.77 to 1.90~\AA\ and are weaker than the B--N bonds.
At the interface, Cu--substrate bonds (Cu--N and Cu--C) are weaker covalent interactions than original B--N and Si--C bonds, as indicated by their longer bond lengths of 1.85--2.02~\AA.
Cu--Cu bonds within the clusters are primarily metallic, with bond lengths of 2.30--2.45~\AA.
A notable exception is found in the single-Cu configuration, where three stable interfacial Si--N bonds form, each measuring about 1.80~\AA.
Bond strengths were additionally evaluated using the Integrated Crystal Orbital Hamilton Population (ICOHP) method~\cite{Nelson2020}, as presented in Fig.~S2.

\subsection*{Chemical reactivity toward gas adsorption}
\label{subsec:ads}
\begin{figure}[htb!]
    \centering
    \includegraphics[width=0.75\columnwidth]{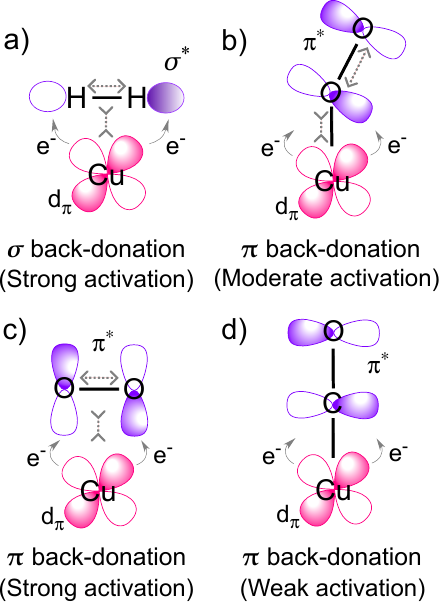}
    \captionsetup{
    justification=justified,
    singlelinecheck=false}
    \caption{Schematic illustration of back-donation from occupied Cu $d_\pi$ orbitals into antibonding molecular orbitals of the adsorbates. (a) Back-donation into the \hh\ $\sigma^*$ orbital, resulting in strong H--H bond activation; (b) back-donation into the \oo\ $\pi^*$ orbital for the bent end-on adsorption configuration, corresponding to moderate O--O activation; (c) back-donation into the \oo\ $\pi^*$ orbital for the side-on adsorption configuration, where interaction with both O atoms leads to strong O--O activation; and (d) back-donation into the \co\ $\pi^*$ orbital, associated with comparatively weak C--O activation. Pink/white lobes represent the Cu $d_\pi$ orbitals, whereas purple/white lobes represent the adsorbate antibonding orbitals. Gray arrows indicate the direction of electron transfer from the Cu states toward the antibonding adsorbate orbitals, while the dashed arrows across the molecular bonds schematically indicate changes in bond strength upon population of these antibonding states.}
    \label{fig:ads-mechanism}
\end{figure}

The interactions of \hh, \nn, \oo, \shh, \co, and \coo\ with Cu$_n$@\vb\ ($1 \leq n \leq 5$) were examined to better characterize the coordination environment of the Cu clusters.
These molecules were selected to span chemically distinct regimes, including electron-withdrawing species (\oo, \co, and \nn), an electron-donating adsorbate (\shh), and highly stable molecules (\coo\ and \nn).
Our choice of cluster sizes was motivated by the evolution of the electronic structure of Cu-decorated systems up to Cu$_5$ (discussed earlier in Section~\ref{subsec:cuadsorption}).

Depending on the specific molecule–substrate complex, adsorption may proceed via a back-donation mechanism.
This process results in the weakening of the intramolecular bonds within the gas species, while spontaneously promotes stronger molecule–substrate bindings.
This phenomenon is elucidated through several examples below.

For side-on ($\eta^2$) \hh\ configuration, the only accessible acceptor orbital is the $\sigma^*$.
In this configuration, the antibonding orbitals efficiently overlap with the Cu $d_\pi$ orbital which promotes substantial $\sigma^*$ population and can lead to significant activation of the H--H bond (Fig.~\ref{fig:ads-mechanism}a).
On the other hand, \oo\ activation is dictated by back-donation into the $\pi^*$ orbitals which is significantly dependent on the adsorption configuration of the molecule.
In the bent end-on ($\eta^1$) configuration, the weak overlap between the Cu $d_\pi$ orbital and the \oo\ $\pi^*$ orbital results in only limited back-donation and moderate O--O activation (Fig.~\ref{fig:ads-mechanism}b).
By contrast, in the $\eta^2$ configuration, \oo\ lies nearly parallel to the cluster surface which promotes effective overlap between the 
Cu $d_\pi$ orbital and both lobes of the \oo\ $\pi^*$ orbital (Fig.~\ref{fig:ads-mechanism}c).
This more effective orbital coupling enhances back-donation and leads to substantial elongation of the O--O bond (i.e. strong \oo\ activation).
In comparison, \co\ binds in a linear geometry with an approximately unchanged C--O bond length, indicating weak activation (Fig.~\ref{fig:ads-mechanism}d).
These considerations provide the necessary basis for the adsorbate-surface interaction analysis that follows.

\begin{figure*}[htb!]
    \centering
    \includegraphics[width=0.8\textwidth]{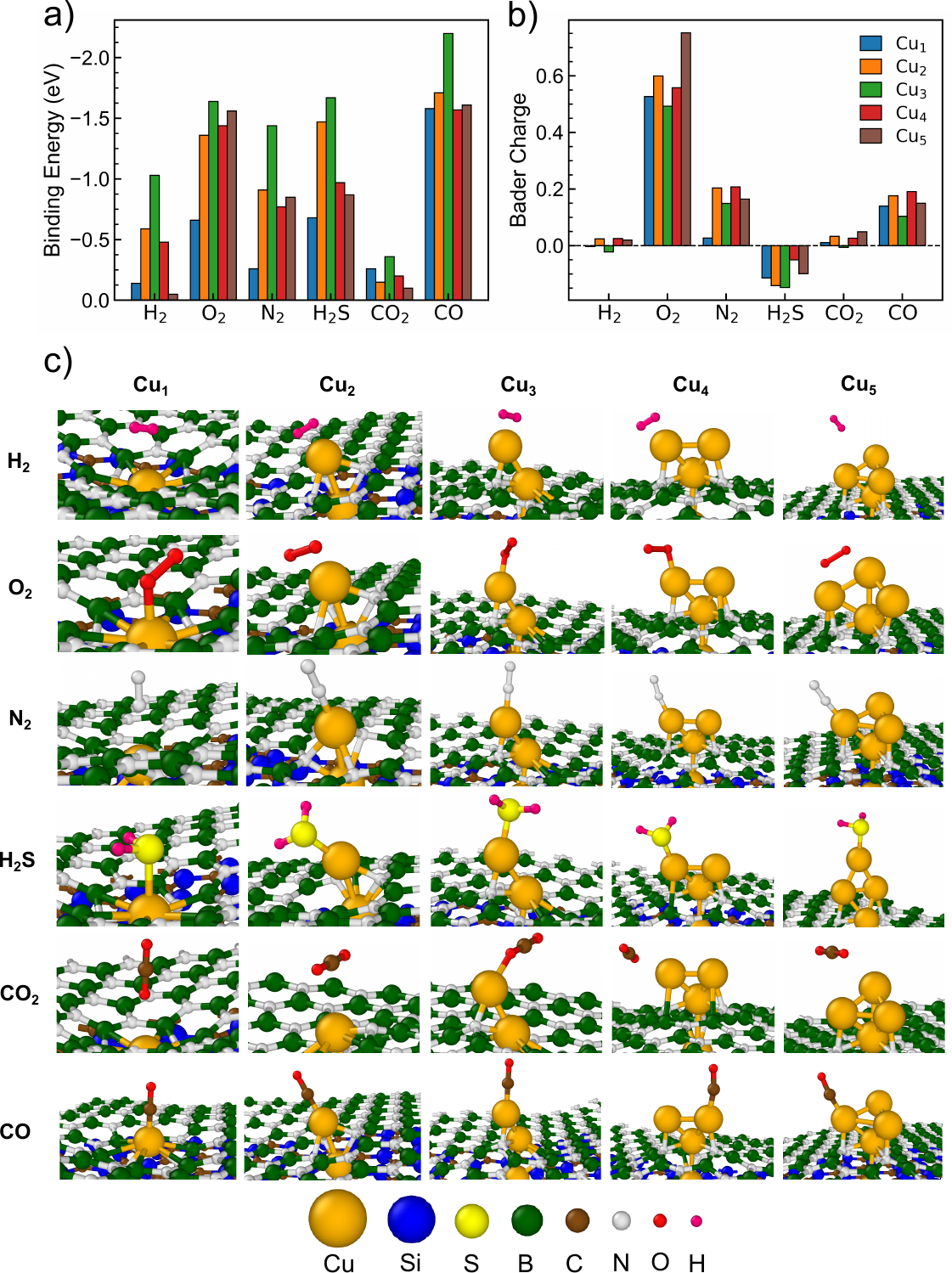}
    \caption{Comparison of (a) adsorption energies, (b) Bader charge transfer ($\Delta q$), and (c) the most stable adsorption geometries of various gas molecules on Cu$_n$@\vb\ ($1 \leq n \leq 5$) in the hBN/SiC heterostructure.
    Positive values of $\Delta q$ indicate charge transfer from the substrate to the adsorbed molecule, whereas negative values indicate charge transfer in the opposite direction. All adsorption calculations were performed using PBE-D3(BJ).}
    \label{fig:adsE-bader}
\end{figure*}
%
The analysis of gas interactions are performed by evaluating the molecular adsorption energies (Fig.~\ref{fig:adsE-bader}a) and net Bader charge transfer ($\Delta q$)~\cite{tang2009grid} between the substrates and adsorbates (Fig.~\ref{fig:adsE-bader}b), together with representative adsorption geometries (Fig.~\ref{fig:adsE-bader}c).
It is important to note that a more negative binding energy indicates stronger adsorption on the surface.
In addition, a positive $\Delta q$ denotes charge transfer from the substrate to the adsorbed molecule, whereas a negative value indicates charge transfer in the opposite direction.

On pristine hBN/SiC, all examined molecules interact weakly with the surface. \co, \hh, \nn, and \shh\ exhibit only weak physisorption, with adsorption energies ranging from approximately $-0.06$ to $-0.19$~eV, depending on the adsorption site (cf. Tables S1--S6).
A similar trend is obtained for the bare \vb-containing hBN/SiC surface, where the strongest interaction is observed for \shh\ but remains limited to $-0.34$~eV.
Indeed, the vacancy site exhibits insufficient affinity toward the adsorption of \co, \oo, and \coo\ (cf. Tables S1--S6).
These results indicate that both pristine and bare \vb-containing hBN/SiC are intrinsically weak adsorption platforms, and require rational modifications to generate chemically accessible adsorption sites.
Anchoring Cu clusters at the \vb\ site, however, substantially modifies the local electronic environment as illustrated in Fig. \ref{fig:fig-str-orb}.

The \hh\ gas indicates an intermediate adsorption regime, only physisorbed on Cu$_1$ and Cu$_5$ with negligible adsorption energies ($-0.14$ and $-0.05$~eV, respectively) and an H--H bond length of $0.75$~\text{\AA}, representing nearly unchanged molecular character.
However, the interactions on Cu$_2$, Cu$_3$, and Cu$_4$, in an $\eta^2$ configuration, are strengthened with adsorption energies of $-0.59$, $-1.03$, and $-0.48$~eV, respectively.
In these instances, the H--H bond elongates to $0.85$~\AA{}.
Specifically, this interaction, with $\eta^2$ orientation of \hh, can be associated with a Kubas-type donation/back-donation mechanism \cite{kubas2001metal,kubas2009hydrogen} which involves $\sigma$-donation from the occupied $\sigma_{\rm{H-H}}$ bonding orbital to Cu together with back-donation from occupied Cu $d$-states to the antibonding $\sigma^{*}_{\rm{H-H}}$ orbital (Fig.~\ref{fig:ads-mechanism}a).

The \oo\ adsorption is highly favorable on all clusters (Cu$_1$--Cu$_5$) with adsorption energies ranging from $-0.66$ to $-1.64$~eV.
The strong binding is associated with substantial Cu-to-\oo\ charge transfer, with $\Delta q$ ranging from $+0.49$ to $+0.75$~$e$.
The large charge transfer may partly arise from $\pi$ back-donation which leads to significant elongation of the O--O bond from $\approx 1.21$~\AA\ in the isolated molecule to $1.30$--$1.39$~\AA\ for molecule-substrate complexes.
However, the degree of activation is linked to the adsorption geometry, Cu$_1$, Cu$_3$, and Cu$_4$ favor a $\eta^1$ oxygen coordination form which possesses a O--O bond of $1.30$--$1.33$~\AA.
In these cases, the orbital alignment and back-donation results in moderate \oo\ activation (Fig.~\ref{fig:ads-mechanism}b).
Over Cu$_2$ and Cu$_5$ clusters, the \oo\ molecule undergoes maximum O--O bond elongation (with bond lengths $\approx$$1.36$ and $1.39$~\AA).
Notably, the $\eta^2$ orientation of \oo\ on Cu$_2$ and Cu$_5$ promotes significantly favorable orbital overlap, thereby enhancing back-donation from Cu d$_\pi$ to the $\pi^*$ orbitals of \oo\ which then leads to a strong \oo\ activation (Fig.~\ref{fig:ads-mechanism}c).
Although Cu$_3$ exhibits the highest adsorption strength, the corresponding \oo\ activation is not comparably favorable.
This indicates that adsorption strength alone is insufficient for evaluating bond activation and back-donation mechanism plays a crucial role.

For \nn\ adsorption, the binding favorability increases from Cu$_1$ to Cu$_3$, reaching $-1.44$~eV on Cu$_3$.
The N--N bond length increases only slightly to $1.12$--$1.13$~\AA, compared with approximately $1.10$~\AA\ for the isolated \nn\ gas.
The Cu--N bond length of $1.78$--$1.82$~\AA\ for Cu$_2$ to Cu$_5$ indicate appreciable metal--molecule interactions, but these interactions induce only limited weakening of the highly stable N$\equiv$N bond.
Thus, \nn\ gas is substantially stabilized on the Cu clusters, particularly on Cu$_3$, while remaining only weakly activated.

The adsorption of \shh\ on the Cu clusters is energetically favorable by $-0.68$, $-1.47$, $-1.67$, $-0.97$, and $-0.87$~eV on Cu$_1$ to Cu$_5$, respectively.
In contrast to the other adsorbates, \shh\ exhibits negative Bader charge values of $-0.11$ to $-0.15$ $e$, indicating net charge transfer from \shh\ to the Cu-decorated substrate.
This opposite charge-transfer direction suggests that the vacancy-stabilized Cu sites can interact with both acceptor-type adsorbates, such as \oo, and donor-type adsorbates, such as \shh.

\begin{figure*}[htbp!]
    \centering
    \includegraphics[width=0.95\textwidth]{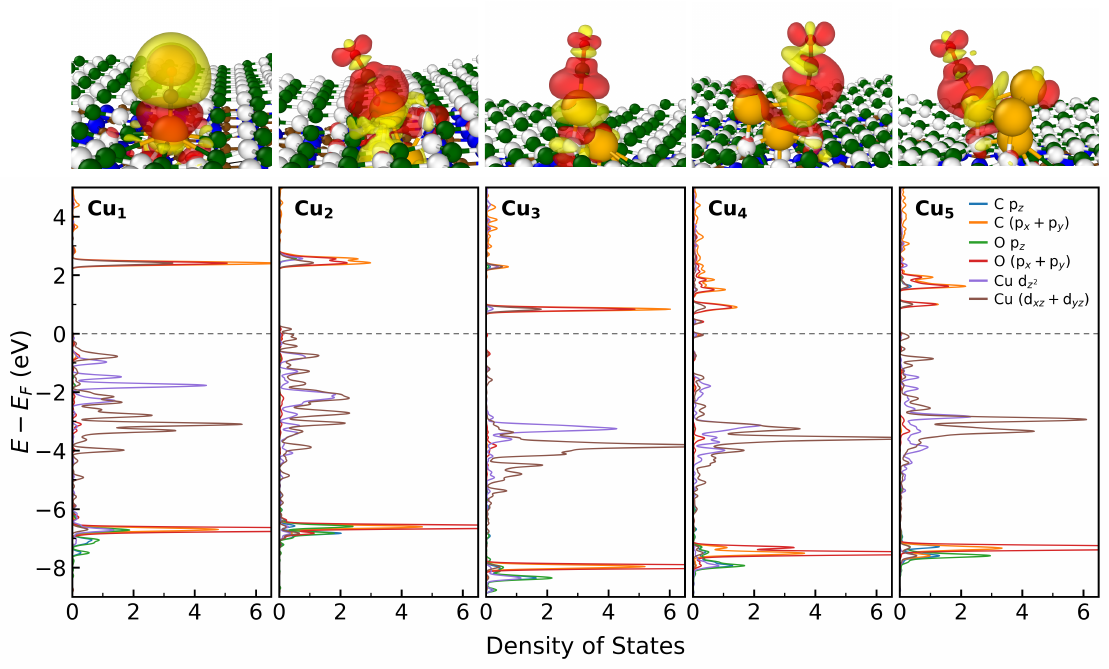}
   \caption{Projected density of states (PDOS) and charge density difference (CDD) for CO adsorption over Cu clusters in hBN/SiC heterostructures. Red and yellow isosurfaces in CDD represent electron accumulation and depletion, respectively. DFT results were obtained using the PBE xc-functional.}
    \label{fig:CO-pdos}
\end{figure*}
%
By contrast, \coo\ remains weakly physisorbed on all the Cu clusters.
Even on Cu$_3$, the adsorption energy is only $-0.36$~eV, with negligible charge transfer and retention of the nearly linear molecular geometry.
Thus, Cu$_1$ to Cu$_5$ complexes exhibit limited affinities for \coo\ adsorption compared with its stronger interactions with \oo, \hh, \nn, and \shh.

The \co\ gas chemisorbs strongly on all Cu clusters, changing from $-1.58$~eV on Cu$_1$ to $-1.71$~eV on Cu$_2$ and reaching $-2.20$~eV on Cu$_3$.
The adsorption strength then decreases to $-1.57$ and $-1.61$~eV on Cu$_4$ and Cu$_5$, respectively.
The corresponding $\Delta q$ values are $+0.14$, $+0.18$, $+0.10$, $+0.19$, and $+0.15$ $e$ for Cu$_1$ to Cu$_5$, respectively.
The strong Cu--\co\ interactions can be interpreted using the Dewar--Chatt--Duncanson model~\cite{dewar1951review, purkayastha2024beryllium}, in which \co\ $\sigma$ donation to Cu is coupled with Cu $\pi$ back-donation into the \co\ $\pi^\ast$ antibonding orbital (Fig. \ref{fig:ads-mechanism}d)~\cite{padilla2008theoretical, ma2023carbon, gameel2018unveiling}.
Across the cluster series, the Cu--C bond length remains nearly constant at $1.77$--$1.79$ \AA, while the C--O bond length changes only slightly to $1.15$--$1.16$ \AA\ compared to $1.13$ \AA\ for the molecular \co\ gas.
Thus, \co\ is strongly chemisorbed through Cu--C bond formation, while the internal C--O bond is only weakly activated.

The Cu--\co\ interaction was further examined using projected densities of states (PDOS) and charge-density-difference (CDD) analyses (Fig.~\ref{fig:CO-pdos}). For Cu$_3$, the \co\ $\pi^*$-derived peaks occur closer to the Fermi level than for Cu$_1$ and Cu$_2$, indicating a nuclearity-dependent change in Cu--\co\ hybridization. This behavior is consistent with an increased contribution from Cu-to-\co\ back-donation, although PDOS peak positions alone are insufficient to quantify the direction or magnitude of charge transfer.
These findings are in line with Refs.~\citenum{gao2023enhancement, amaya2020adsorption}.

The CDD maps show electron accumulation in the Cu--C bonding region and electron depletion around neighboring Cu atoms, particularly for Cu$_2$--Cu$_5$. This redistribution supports the formation of a Cu--C bond is still, by itself, insufficient to fully confirm the $\sigma$-donation
and $\pi$-back-donation contributions. However, considered together with the PDOS, Bader charge transfer, and bond lengths, the results are consistent with contributions from both interaction mechanisms. The limited C--O elongation indicates that the internal C--O bond remains only weakly
activated.

Overall, the adsorption results reveal a clear nuclearity dependence across the Cu$_1$--Cu$_5$ series.
The Cu$_3$ complex emerges as the adsorption-strength optimum for all examined molecules, giving the most favorable adsorption energies for \co, \hh, \nn, \oo, \shh, and \coo, with values of $-2.20$, $-1.03$, $-1.44$, $-1.64$, $-1.67$, and $-0.36$~eV, respectively.
This maximum arises from the specific asymmetric geometry of Cu$_3$@\vb, in which two Cu atoms are stabilized within the vacancy and in the interlayer region, while the third Cu atom remains exposed and under-coordinated.
This exposed Cu site provides favorable orbital overlap with incoming adsorbates, whereas the embedded Cu atoms contribute to stabilizing the hBN and SiC layers through interfacial bonding and charge redistribution.
The decrease in adsorption strength for Cu$_4$ and Cu$_5$ indicates that further Cu incorporation partially reduces the unique exposed-site character of Cu$_3$.

The PDOS and CDD results for the other investigated gases are presented in Figs. S3–S7.
The relative and adsorption energies of all evaluated adsorption configurations are summarized in Tables S1–S6, and the corresponding bond lengths are provided in Table S7.

\section*{Discussion}
In this work, we have established a computational framework for understanding how vacancy engineering in a semiconducting hBN/SiC heterostructure, as a novel substrate, can be used to stabilize Cu single atoms and few-atom clusters.
Single metal atoms on weakly interacting surfaces are generally likely to result in migration and aggregation, which can reduce the efficiency of atom utilization and complicate the precise identification of the active site.
Our results show that \vb\ in the hBN layer provides an effective anchoring motif for Cu, linking defect-assisted metal stabilization to the formation of well-defined, electronically tunable active centers.
At low Cu loading, the first Cu atom preferentially occupies the interlayer region near \vb, where it coordinates with under-coordinated N atoms and partially compensates the local electronic deficiency associated with the vacancy \cite{zhu2017taming}.
This metal--defect interaction weakens the native interlayer bonding between the N atoms surrounding \vb\ and the underlying Si/C atoms of the SiC monolayer, indicating a redistribution of bonding at the hBN/SiC interface.
With increasing Cu loading, the local coordination environment reorganizes and Cu--Cu interactions emerge, leading to anchored Cu$_n$ motifs rather than isolated, less-substrate-supported metal atoms.
These metal--defect and metal--metal interactions stabilize highly dispersed Cu species while enabling charge redistribution between the substrate and the Cu cluster.

The adsorption calculations reveal that this structural stabilization is directly coupled to chemical activity.
Pristine hBN/SiC and bare \vb-containing hBN/SiC interact only weakly with the examined gas molecules, indicating that the vacancy itself acts primarily as an anchoring and interfacial stabilization site rather than as a strongly reactive center.
In contrast, Cu incorporation substantially modifies the local electronic structure and creates under-coordinated metal sites capable of stronger molecule--surface interactions.
This behavior is consistent with experimental evidence showing that electron transfer between transition-metal species and \vb-containing hBN supports enhances the reactivity of the resulting heterostructures~\cite{zhu2017taming}.
Overall, changes in the size of the Cu clusters lead to a nonlinear relationship between cluster size and adsorption strength.
Indeed, the incorporation of the second Cu atom drives the first Cu atom deeper into the interlayer region and both Cu atoms together contribute to compensating the charge deficiency associated with the \vb~defect. Consequently, an active site is formed, although this configuration remains less favorable than the optimal case since the Cu on the top is still not strongly under-coordinated.
Together, this Cu dimer establishes a suitable structural foundation for the third Cu atom, which is placed at the top of the cluster. However, the third Cu atom contributes only weakly to compensating the charge deficiency of \vb\ and remains highly under-coordinated. Consequently, Cu$_3$ emerges as the optimal motif among the considered systems, offering a balance between stabilization within the \vb\ defect, electronic flexibility, and accessible under-coordinated Cu which acts as single active site on the surface. This balance allows the cluster to reorganize charge upon molecular adsorption while maintaining stable cluster--hBN/SiC binding.
In contrast, smaller and larger Cu motifs lack the same balance of structural stability, electronic flexibility, and chemical responsiveness, emphasizing the importance of cluster nuclearity in regulating adsorption behavior at defect-engineered hBN/SiC interfaces.
These findings suggest a design principle in which vacancy-stabilized metal clusters can be used to tune the reactivity of otherwise chemically inert two-dimensional heterostructures through controlled metal loading and interfacial charge redistribution.

\section*{Methods}
\label{sect:methods}

All DFT calculations were performed using the Vienna \emph{Ab initio} Simulation Package (VASP)~\cite{kres1} within the projector augmented-wave (PAW) formalism~\cite{PAW1994,kresse1999ultrasoft}.
Exchange--correlation effects were described using the Perdew--Burke--Ernzerhof (PBE) functional~\cite{Perdew1996_PRL}, while selected electronic-structure calculations were carried out with the screened hybrid Heyd--Scuseria--Ernzerhof functional, HSE03~\cite{Heyd2003_JCP}, to obtain improved orbital energies and band gaps.
As discussed in our previous study~\cite{Hashemi2026_PRM}, the HSE03 functional yielded electronic band gaps for hBN and SiC monolayers that are in close agreement with the corresponding experimental values.
PBE was used for structural relaxation, relative energetics, and gas-adsorption calculations, providing an efficient description of the large defect-engineered hBN/SiC systems considered here.
Long-range dispersion interactions were included through the DFT-D3 correction with Becke--Johnson damping~\cite{Grimme2011_JCC}.
A plane-wave cutoff energy of 500 eV was used throughout, and the electronic self-consistency criterion was set to $10^{-6}$\,eV.
All structures were fully relaxed until the residual forces on each atom were below $0.01$ eV/\AA. 
It must be noted that PBE-D3 may predict stronger adsorption for gases such as CO and O$_2$ at exposed metal sites as previously observed for CO adsorption at vacancy-exposed Mo sites in two-dimensional MoSe$_2$ \cite{bombin2024scattering}.
Therefore, the absolute adsorption energies should be interpreted with consideration of this functional dependence, while comparisons among the Cu$_n$@\vb–hBN/SiC systems remain qualitatively meaningful because all systems were treated consistently at the same level of theory.

The model system consisted of a bilayer hBN/SiC heterostructure containing 100 B, 100 N, 64 Si, and 64 C atoms in the pristine supercell.
A \vb\ defect was introduced by removing a single B atom.
Upon structural relaxation, interlayer chemical bonds form around the vacancy site, locally converting the van der Waals interface into a chemically bonded structure, as detailed in our previous work ~\cite{Hashemi2026_PRM}.
The interface was constructed by matching a $5\times5\times1$ hBN supercell with a $4\times4\times1$ SiC supercell, corresponding to one quarter of the full supercell used in our study, while maintaining minimal lattice mismatch.
A vacuum spacing of 25~\AA\ was applied along the out-of-plane direction, and the Brillouin zone was sampled at the $\Gamma$ point.

To investigate Cu adsorption and trapping, we performed MLMD simulations using a neuroevolution potential (NEP)~\cite{song2024general}.
The simulation cell was expanded to 117.5~\AA\ and 67.8~\AA\ along the $x$ and $y$ directions, respectively, giving an in-plane lattice mismatch of 0.026\%.
A vacuum spacing of 45~\AA\ was applied along the $z$-direction.
Ten \vb\ defects were randomly introduced into the hBN layer, corresponding to a defect density of $3.44\times10^{3}$~ppm, within the experimentally reported range of $5.4\times10^{2}$--$1.8\times10^{4}$~ppm~\cite{gong2023coherent}.
It should be noted that this experimental range was reported for spin defects in monolayer hBN rather than in hBN/SiC heterostructures.
Cu atoms were then randomly placed 3~\AA\ above the defective hBN surface, with the number of Cu atoms varied from 5 to 60 to obtain Cu/\vb\ ratios of 1, 2, 4, and 6.
For each composition, ten independent simulations were performed, and the averaged results are reported.
In each trial, a distinct defective configuration was generated.
The resulting defective Cu-decorated hBN/SiC interfaces were equilibrated in the canonical ($NVT$) ensemble for 5~ns, allowing Cu atoms to diffuse, interact with vacancy sites, and form stabilized configurations. These
simulations were carried out at 300~K.
All MLMD simulations were performed with the GPUMD package~\cite{xu2025gpumd} using a timestep of 0.5~fs.
The NEP model was trained using the PBE-level dataset developed our recent work~\cite{Hashemi2026_PRM}, which was further extended in this work to include isolated Cu clusters and Cu clusters supported on pristine hBN/SiC.

The binding energy of the gas-phase adsorbate (X) on the surface is defined as:
\begin{equation}
    E_{\mathrm{bind}} = E_{\mathrm{X@S}} - E_{\mathrm{S}} - E_{\mathrm{X}},
    \label{eq:bindenergy}
\end{equation}
where $E_{\mathrm{X@S}}$ is the total energy of the fully relaxed adsorption complex, $E_{\mathrm{S}}$ is the total energy of the corresponding relaxed substrate without the adsorbate, and $E_{\mathrm{X}}$ is the total energy of the gas species in vacuum. 
Here, S denotes either the defective hBN/SiC substrate or a Cu-decorated Cu$_n$@\vb--hBN/SiC surface, depending on the adsorption site considered.
This definition can be extended to include both Cu adsorption and Cu-cluster formation, thereby enabling a fundamental understanding of the thermodynamic driving force behind Cu mono-dispersion, as follows:
\begin{equation}
    E_{\mathrm{f}} = E_{\mathrm{complex}} - E_{\rm{S}} - N_{\rm{atom}} E_{\mathrm{atom}},
    \label{eq:cu_ads_n}
\end{equation}
where $E_{\mathrm{complex}}$, $E_{{\rm{S}}}$, and $E_{\mathrm{atom}}$ are the total energies of the relaxed defective system containing $N_{\rm{atom}}$ Cu atoms, the \vb-containing hBN/SiC substrate, and an isolated Cu atom in vacuum, respectively.
It is worth noting that the present calculations consider isolated gas molecules and adsorption on the same substrate without explicit entropic corrections. Although entropy contributions may affect the absolute adsorption free energies, the calculated adsorption energies are expected to provide a reasonable qualitative comparison among the gases.
Nevertheless, entropy may influence their relative ordering, particularly when the differences in adsorption energy are small.

The minimum-energy pathways were determined employing the climbing-image nudged elastic band (CI-NEB) method, as implemented in the Atomic Simulation Environment (ASE) Python library~\cite{HjorthLarsen_2017} to validate the MLMD potential.
Each pathway was represented by 15 images, including the endpoints, initialized using image-dependent pair-potential interpolation with the minimum-image convention.
The improved-tangent formulation and a spring constant of 0.20~eV/\AA$^{2}$ were employed.
The band was optimized using the FIRE algorithm through successive regular- and climbing-image stages with progressively tighter settings, until the maximum NEB force was below 0.05~eV/\AA.
All interatomic interactions were described using the trained NEP model, as implemented in the GPUMD software and imported via the CALORINE~\cite{Lindgren2024} package.

\section*{Data Availability}
The data that support the findings of this study are available upon request.

\section*{Acknowledgments}
\label{sect:ack}
We are grateful to  CSC--IT Center for Science Ltd. and Aalto Science-IT project for generous grants of computer time.
R. K., X. C., and T.A-N. acknowledge support from the Research Council of Finland grant no. 370057, and
A. H. and M. K. were supported by the Research Council of Finland Flagship of Advanced Mathematics for Sensing Imaging and Modelling, grant 358944 and by Foundation PS.

\bibliography{main.bib}

@article{thakur2024accelerating,
  title={Accelerating 2D materials discovery},
  author={Thakur, Anupma and Anasori, Babak},
  journal={Science},
  volume={383},
  number={6688},
  pages={1182--1183},
  year={2024},
  doi = {10.1126/science.ado411},
  publisher={American Association for the Advancement of Science}
}

@article{katiyar20232d,
  title={2D materials in flexible electronics: recent advances and future prospectives},
  author={Katiyar, Ajit Kumar and Hoang, Anh Tuan and Xu, Duo and Hong, Juyeong and Kim, Beom Jin and Ji, Seunghyeon and Ahn, Jong-Hyun},
  journal={Chemical Reviews},
  volume={124},
  number={2},
  pages={318--419},
  year={2024},
  doi={10.1021/acs.chemrev.3c00302},
  publisher={ACS Publications}
}

@article{liang2022progress,
  title={The progress and outlook of metal single-atom-site catalysis},
  author={Liang, Xiao and Fu, Ninghua and Yao, Shuangchao and Li, Zhi and Li, Yadong},
  journal={J. Am. Chem. Soc.},
  volume={144},
  number={40},
  pages={18155--18174},
  year={2022},
  doi = {10.1021/jacs.1c12642},
  publisher={ACS Publications}
}

@article{joudi2025two,
  title={Two-Dimensional One-Atom-Thick Gold Grown on Defect-Engineered Graphene},
  author={Joudi, Wael and Ghaderzadeh, Sadegh and Trentino, Alberto and Mizohata, Kenichiro and Mustonen, Kimmo and Besley, Elena and Kotakoski, Jani and {\AA}hlgren, E Harriet},
  journal={ACS Nano},
  volume = {19},
  pages = {22032--22043},
  doi = {10.1021/acsnano.5c01538},
  year={2025},
  publisher={ACS Publications}
}

@article{song2024general,
  title={General-purpose machine-learned potential for 16 elemental metals and their alloys},
  author={Song, Keke and Zhao, Rui and Liu, Jiahui and Wang, Yanzhou and Lindgren, Eric and Wang, Yong and Chen, Shunda and Xu, Ke and Liang, Ting and Ying, Penghua and others},
  journal={Nat. Commun.},
  volume={15},
  number={1},
  pages={10208},
  year={2024},
  doi = {10.1038/s41467-024-54554-x},
  publisher={Nature Publishing Group UK London}
}

@article{kres1,
  title = {Ab Initio Molecular Dynamics For Liquid Metals},
  author = {Kresse, G. and Hafner, J.},
  journal = {Phys. Rev. B},
  volume = {47},
  issue = {1},
  pages = {558--561},
  numpages = {0},
  year = {1993},
  month = {Jan},
  publisher = {American Physical Society},
  doi = {10.1103/PhysRevB.47.558},
  url = {http://link.aps.org/doi/10.1103/PhysRevB.47.558}
}

@article{PAW1994,
  title = {Projector augmented-wave method},
  author = {Bl\"ochl, P. E.},
  journal = {Phys. Rev. B},
  volume = {50},
  issue = {24},
  pages = {17953--17979},
  numpages = {0},
  year = {1994},
  month = {Dec},
  publisher = {American Physical Society},
  doi = {10.1103/PhysRevB.50.17953},
  url = {https://link.aps.org/doi/10.1103/PhysRevB.50.17953}
}

@article{Perdew1996_PRL,
  title = {Generalized Gradient Approximation Made Simple},
  author = {Perdew, John P. and Burke, Kieron and Ernzerhof, Matthias},
  journal = {Phys. Rev. Lett.},
  volume = {77},
  issue = {18},
  pages = {3865--3868},
  numpages = {0},
  year = {1996},
  month = {Oct},
  publisher = {American Physical Society},
  doi = {10.1103/PhysRevLett.77.3865},
  url = {https://link.aps.org/doi/10.1103/PhysRevLett.77.3865}
}

@article{Heyd2003_JCP,
    author = {Heyd, Jochen and Scuseria, Gustavo E. and Ernzerhof, Matthias},
    title = {Hybrid functionals based on a screened Coulomb potential},
    journal = {J. Chem. Phys.},
    volume = {118},
    number = {18},
    pages = {8207--8215},
    year = {2003},
    month = {05},
    issn = {0021-9606},
    doi = {10.1063/1.1564060}
}

@article{Grimme2011_JCC,
author = {Grimme, Stefan and Ehrlich, Stephan and Goerigk, Lars},
title = {Effect of the damping function in dispersion corrected density functional theory},
journal = {J. Comput. Chem.},
volume = {32},
number = {7},
pages = {1456--1465},
doi = {10.1002/jcc.21759},
year = {2011}
}

@article{gong2023coherent,
  title={Coherent dynamics of strongly interacting electronic spin defects in hexagonal boron nitride},
  author={Gong, Ruotian and He, Guanghui and Gao, Xingyu and Ju, Peng and Liu, Zhongyuan and Ye, Bingtian and Henriksen, Erik A and Li, Tongcang and Zu, Chong},
  journal={Nat. Commun.},
  volume={14},
  number={1},
  pages={3299},
  year={2023},
  doi = {10.1038/s41467-023-39115-y},
  publisher={Nature Publishing Group UK London}
}

@article{wu2023mdapy,
  title={mdapy: A flexible and efficient analysis software for molecular dynamics simulations},
  author={Wu, Yong-Chao and Shao, Jian-Li},
  journal={Comput. Phys. Comm.},
  volume={290},
  pages={108764},
  year={2023},
  doi = {10.1016/j.cpc.2023.108764},
  publisher={Elsevier}
}

@article{lee2003semiempirical,
  title={Semiempirical atomic potentials for the fcc metals {Cu, Ag, Au, Ni, Pd, Pt, Al, and Pb} based on first and second nearest-neighbor modified embedded atom method},
  author={Lee, Byeong-Joo and Shim, Jae-Hyeok and Baskes, MI},
  journal={Phys. Rev. B},
  volume={68},
  number={14},
  pages={144112},
  year={2003},
  doi = {10.1103/PhysRevB.68.144112},
  publisher={APS}
}

@article{xu2025gpumd,
  title={{GPUMD} 4.0: a high-performance molecular dynamics package for versatile materials simulations with machine-learned potentials},
  author={Xu, Ke and Bu, Hekai and Pan, Shuning and Lindgren, Eric and Wu, Yongchao and Wang, Yong and Liu, Jiahui and Song, Keke and Xu, Bin and Li, Yifan and others},
  journal={Materials Genome Engineering Advances},
  volume={3},
  number={3},
  pages={e70028},
  year={2025},
  doi = {10.1002/mgea.70028},
  publisher={Wiley Online Library}
}

@article{Hashemi2026_PRM,
  title = {Stabilization of {hBN/SiC} heterostructures with vacancies and transition-metal atoms},
  author = {Hashemi, Arsalan and Cherati, Nima Ghafari and Ghaderzadeh, Sadegh and Wang, Yanzhou and Ghorbani-Asl, Mahdi and Ala-Nissila, Tapio},
  journal = {Phys. Rev. Mater.},
  volume = {10},
  issue = {4},
  pages = {045801},
  numpages = {12},
  year = {2026},
  month = {Apr},
  doi = {10.1103/lymz-nlbf},
  publisher = {American Physical Society},
  url = {https://link.aps.org/doi/10.1103/lymz-nlbf}
}

@article{padilla2008theoretical,
  title={Theoretical study of the adsorption of carbon monoxide on small copper clusters},
  author={Padilla-Campos, Luis},
  journal={Journal of Molecular Structure: THEOCHEM},
  volume={851},
  number={1-3},
  pages={15--21},
  year={2008},
  doi = {10.1016/j.theochem.2007.10.027},
  publisher={Elsevier}
}

@article{ma2023carbon,
  title={Carbon monoxide separation: past, present and future},
  author={Ma, Xiaozhou and Albertsma, Jelco and Gabriels, Dieke and Horst, Rens and Polat, Sevgi and Snoeks, Casper and Kapteijn, Freek and Eral, H{\"u}seyin Burak and Vermaas, David A and Mei, Bastian and others},
  journal={Chemical Society Reviews},
  volume={52},
  number={11},
  pages={3741--3777},
  year={2023},
  doi = {10.1039/d3cs00147d},
  publisher={Royal Society of Chemistry}
}

@article{gameel2018unveiling,
  title={Unveiling {CO} adsorption on {Cu} surfaces: new insights from molecular orbital principles},
  author={Gameel, Kareem M and Sharafeldin, Icell M and Abourayya, Amr U and Biby, Ahmed H and Allam, Nageh K},
  journal={Phys. Chem. Chem. Phys.},
  volume={20},
  number={40},
  pages={25892--25900},
  doi = {10.1039/c8cp04253e},
  year={2018},
  publisher={Royal Society of Chemistry}
}

@article{gao2023enhancement,
  title={Enhancement of {CO} adsorption energy on defective graphene-supported {Cu}$_{13}$ cluster and prediction with an induction energy model},
  author={Gao, Delu and Rao, Shenyan and Li, Yueru and Liu, Naigui and Wang, Dunyou},
  journal={Applied Surface Science},
  volume={615},
  pages={156368},
  year={2023},
  doi = {10.1016/j.apsusc.2023.156368},
  publisher={Elsevier}
}

@article{amaya2020adsorption,
  title={Adsorption and dissociation of {CO} on metal clusters},
  author={Amaya-Roncancio, Sebastian and Reinaudi, Luis and Gimenez, M Cecilia},
  journal={Materials Today Communications},
  volume={24},
  pages={101158},
  year={2020},
  doi = {10.1016/j.mtcomm.2020.101158},
  publisher={Elsevier}
}

@article{kubas2001metal,
  title={{Metal--dihydrogen and $\sigma$-bond coordination: the consummate extension of the Dewar--Chatt--Duncanson model for metal--olefin $\pi$ bonding}},
  author={Kubas, Gregory J},
  journal={J. Organomet. Chem.},
  volume={635},
  number={1-2},
  pages={37--68},
  year={2001},
  doi = {10.1016/S0022-328X(01)01066-X},
  publisher={Elsevier}
}

@article{kubas2009hydrogen,
  title={Hydrogen activation on organometallic complexes and {H$_2$} production, utilization, and storage for future energy},
  author={Kubas, Gregory J},
  journal={J. Organomet. Chem.},
  volume={694},
  number={17},
  pages={2648--2653},
  year={2009},
  doi = {10.1016/j.jorganchem.2009.05.027},
  publisher={Elsevier}
}

@article{dewar1951review,
  title={A review of the pi-complex theory},
  author={Dewar, JS},
  journal={Bulletin de la Societe Chimique de France},
  volume={18},
  number={3-4},
  pages={C71--C79},
  year={1951},
  publisher={EDITIONS SCIENTIFIQUES MEDICALES ELSEVIER 23 RUE LINOIS, 75724 PARIS CEDEX~…}
}

@article{purkayastha2024beryllium,
  title={Beryllium carbonyl {Be(CO)$_n$ ($n$ = 1--4) complex: a p-orbital analogy of Dewar--Chatt--Duncanson model}},
  author={Purkayastha, Siddhartha K and Rohman, Shahnaz S and Parameswaran, Pattiyil and Guha, Ankur K},
  journal={Phys. Chem. Chem. Phys.},
  volume={26},
  number={16},
  pages={12573--12579},
  year={2024},
  doi = {10.1039/d4cp00908h},
  publisher={Royal Society of Chemistry}
}

@article{Kohlrausch2025_AdvSci,
author = {Kohlrausch, Emerson C. and Ghaderzadeh, Sadegh and Aliev, Gazi N. and Popov, Ilya and Saad, Fatmah and Alharbi, Eman and Ramasse, Quentin M. and Rance, Graham A. and Danaie, Mohsen and Thangamuthu, Madasamy and Young, Mathew and Plummer, Richard and Morgan, David J. and Theis, Wolfgang and Besley, Elena and Khlobystov, Andrei N. and Alves Fernandes, Jesum},
title = {One-Size-Fits-All: {A} Universal Binding Site for Single-Layer Metal Cluster Self-Assembly},
journal = {Adv. Sci.},
volume = {12},
number = {37},
pages = {e08034},
doi = {10.1002/advs.202508034},
url = {https://advanced.onlinelibrary.wiley.com/doi/abs/10.1002/advs.202508034},
year = {2025}
}

@article{Popov_nanolet_2023,
author = {Popov, Ilya and Ghaderzadeh, Sadegh and Kohlrausch, Emerson C. and Norman, Luke T. and Slater, Thomas J. A. and Aliev, Gazi N. and Alhabeadi, Hanan and Kaplan, Andre and Theis, Wolfgang and Khlobystov, Andrei N. and Fernandes, Jesum Alves and Besley, Elena},
title = {Chemical Kinetics of Metal Single Atom and Nanocluster Formation on Surfaces: {An} Example of {Pt} on Hexagonal Boron Nitride},
journal = {Nano Lett.},
volume = {23},
number = {17},
pages = {8006--8012},
year = {2023},
note ={PMID: 37594260},
doi = {10.1021/acs.nanolett.3c01968},
url = {https://doi.org/10.1021/acs.nanolett.3c01968},
}

@article{Ghaderzadeh_acs_2026,
author = {Ghaderzadeh, Sadegh and Popov, Ilya and Theis, Wolfgang and Alves Fernandes, Jesum and Khlobystov, Andrei N. and Besley, Elena},
title = {Platinum Atoms Dynamics on the Surface of Hexagonal Boron Nitride Containing Vacancy Defects},
journal = {ACS Applied Materials \& Interfaces},
volume = {18},
number = {5},
pages = {9216--9224},
year = {2026},
note ={PMID: 41592215},
doi = {10.1021/acsami.5c22977},
url = {https://doi.org/10.1021/acsami.5c22977},
}

@article{chen2023prediction,
  title={Prediction of three-metal cluster catalysts on two-dimensional {W$_2$N$_3$} support with integrated descriptors for electrocatalytic nitrogen reduction},
  author={Chen, Siyu and Gao, Yongqi and Wang, Wugang and Prezhdo, Oleg V and Xu, Lai},
  journal={ACS Nano},
  volume={17},
  number={2},
  pages={1522--1532},
  year={2023},
  doi = {10.1021/acsnano.2c10607},
  publisher={ACS Publications}
}

@article{zheng2024growing,
  title={Growing highly ordered Pt and Mn bimetallic single atomic layers over graphdiyne},
  author={Zheng, Zhiqiang and Qi, Lu and Luan, Xiaoyu and Zhao, Shuya and Xue, Yurui and Li, Yuliang},
  journal={Nat. Commun.},
  volume={15},
  number={1},
  pages={7331},
  year={2024},
  doi = {10.1038/s41467-024-51687-x},
  publisher={Nature Publishing Group UK London}
}

@article{liu2022dual,
  title={Dual transition metal atoms embedded in {N}-doped graphene for electrochemical nitrogen fixation under ambient conditions},
  author={Liu, Yi and Song, Bingyi and Huang, Chun-Xiang and Yang, Li-Ming},
  journal={J. Mat. Chem. A},
  volume={10},
  number={25},
  pages={13527--13543},
  year={2022},
  doi = {10.1039/d1ta11024a},
  publisher={Royal Society of Chemistry}
}

@article{bord2023atomistic,
  title={An atomistic view of platinum cluster growth on pristine and defective graphene supports},
  author={Bord, Julia and Kirchhoff, Bj{\"o}rn and Baldofski, Matthias and Jung, Christoph and Jacob, Timo},
  journal={Small},
  volume={19},
  number={10},
  pages={2207484},
  year={2023},
  doi = {10.1002/smll.202207484},
  publisher={Wiley Online Library}
}

@article{lang2019non,
  title={Non defect-stabilized thermally stable single-atom catalyst},
  author={Lang, Rui and Xi, Wei and Liu, Jin-Cheng and Cui, Yi-Tao and Li, Tianbo and Lee, Adam Fraser and Chen, Fang and Chen, Yang and Li, Lei and Li, Lin and others},
  journal={Nat. Commun.},
  volume={10},
  number={1},
  pages={234},
  year={2019},
  doi = {10.1038/s41467-018-08136-3},
  publisher={Nature Publishing Group UK London}
}

@article{liu2024understanding,
  title={Understanding the dynamic aggregation in single-atom catalysis},
  author={Liu, Laihao and Chen, Tiankai and Chen, Zhongxin},
  journal={Adv. Sci.},
  volume={11},
  number={13},
  pages={2308046},
  year={2024},
  doi = {10.1002/advs.202308046},
  publisher={Wiley Online Library}
}

@article{Herrera-Reinoza2021_CM,
author = {Herrera-Reinoza, Nataly and dos Santos, Alisson Ceccatto and de Lima, Luis Henrique and Landers, Richard and de Siervo, Abner},
title = {Atomically Precise Bottom-Up Synthesis of {h-BNC: Graphene Doped with h-BN Nanoclusters}},
journal = {Chemistry of Materials},
volume = {33},
number = {8},
pages = {2871--2882},
year = {2021},
doi = {10.1021/acs.chemmater.1c00081},
url = {https://doi.org/10.1021/acs.chemmater.1c00081}
}

@article{thomas2024colloidal,
  title={Colloidal {2D} layered {SiC} quantum dots from a liquid precursor: surface passivation, bright photoluminescence, and planar self-assembly},
  author={Thomas, Salim A and Alharthi, Naif S and Petersen, Reed J and Aldrees, Ahmed and Tani, Sakurako and Anderson, Kenneth J and Granlie, Joseph and Pringle, Todd A and Payne, Scott A and Choi, Yongki and others},
  journal={ACS Nano},
  volume={18},
  number={39},
  pages={26848--26857},
  year={2024},
  doi = {10.1021/acsnano.4c08052},
  publisher={ACS Publications}
}

@article{polley2023bottom,
  title={Bottom-up growth of monolayer honeycomb {SiC}},
  author={Polley, CM and Fedderwitz, H and Balasubramanian, T and Zakharov, AA and Yakimova, Rositsa and B{\"a}cke, O and Ekman, J and Dash, SP and Kubatkin, S and Lara-Avila, S},
  journal={Phys. Rev. Lett.},
  volume={130},
  number={7},
  pages={076203},
  year={2023},
  doi = {10.1103/PhysRevLett.130.076203},
  publisher={APS}
}

@article{qi2023modulating,
  title={Modulating electronic structures of iron clusters through orbital rehybridization by adjacent single copper sites for efficient oxygen reduction},
  author={Qi, Chunhong and Yang, Haoyu and Sun, Ziqi and Wang, Haifeng and Xu, Na and Zhu, Guihua and Wang, Lianjun and Jiang, Wan and Yu, Xiqian and Li, Xiaopeng and others},
  journal={Angewandte Chemie International Edition},
  volume={62},
  number={39},
  pages={e202308344},
  year={2023},
  doi = {10.1002/ange.202308344Digital Object Identifier (DOI)},
  publisher={Wiley Online Library}
}

@article{yang2022copper,
  title={Copper-involved highly efficient oxygen reduction reaction in both alkaline and acidic media},
  author={Yang, Zehua and Jiang, Kaiyue and Tong, Gangsheng and Ke, Changchun and Wu, Haofei and Liu, Pan and Zhang, Jichao and Ji, Huiping and Zhu, Jinhui and Lu, Chenbao and others},
  journal={Chemical Engineering Journal},
  volume={437},
  pages={135377},
  year={2022},
  doi = {10.1016/j.cej.2022.135377},
  publisher={Elsevier}
}

@article{lin2012light,
  title={Light-emitting two-dimensional ultrathin silicon carbide},
  author={Lin, SS},
  journal={J. Phys. Chem. C},
  volume={116},
  number={6},
  pages={3951--3955},
  year={2012},
  doi = {10.1021/jp210536m},
  publisher={ACS Publications}
}

@article{lin2022boron,
  title={Boron nitride on {SiC} (0001)},
  author={Lin, You-Ron and Franke, Markus and Parhizkar, Shayan and Raths, Miriam and Wen-zhe Yu, Victor and Lee, Tien-Lin and Soubatch, Serguei and Blum, Volker and Tautz, F Stefan and Kumpf, Christian and others},
  journal={Physical Review Materials},
  volume={6},
  number={6},
  pages={064002},
  doi = {10.1103/PhysRevMaterials.6.064002},
  year={2022},
  publisher={APS}
}

@article{zhu2017taming,
  title={Taming interfacial electronic properties of platinum nanoparticles on vacancy-abundant boron nitride nanosheets for enhanced catalysis},
  author={Zhu, Wenshuai and Wu, Zili and Foo, Guo Shiou and Gao, Xiang and Zhou, Mingxia and Liu, Bin and Veith, Gabriel M and Wu, Peiwen and Browning, Katie L and Lee, Ho Nyung and others},
  journal={Nat. Commun.},
  volume={8},
  number={1},
  pages={15291},
  year={2017},
  doi = {10.1038/ncomms15291},
  publisher={Nature Publishing Group UK London}
}

@article{zeng2026liquid,
  title={Liquid metal dispersed single-atom catalyst with high-temperature stability},
  author={Zeng, Ziyue and Wang, Chenyang and Sun, Mingjun and Liu, Dong and Zhang, Donghong and He, Shiyi and Liang, Zijia and Ma, Yushen and Zhang, Yile and Li, Ling and others},
  journal={Nat. Commun.},
  volume={17},
  number={1},
  pages={3918},
  year={2026},
  doi = {10.1038/s41467-026-70476-2},
  publisher={Nature Publishing Group}
}

@article{Nelson2020,
author = {Nelson, Ryky and Ertural, Christina and George, Janine and Deringer, Volker L. and Hautier, Geoffroy and Dronskowski, Richard},
title = {{LOBSTER: Local} orbital projections, atomic charges, and chemical-bonding analysis from projector-augmented-wave-based density-functional theory},
journal = {J. Comput. Chem.},
volume = {41},
number = {21},
pages = {1931--1940},
doi = {10.1002/jcc.26353},
url = {https://onlinelibrary.wiley.com/doi/abs/10.1002/jcc.26353},
year = {2020}
}

@article{bombin2024scattering,
  title={{Scattering of {CO} from vacant-{MoSe$_2$} with {O} adsorbates: Is {CO$_2$} formed?}},
  author={Bomb{\'\i}n, Ra{\'u}l and D{\'i}ez Mui{\~n}o, Ricardo and Juaristi, J Inaki and Alducin, Maite},
  journal={J. Phys. Chem. C},
  volume={128},
  number={46},
  pages={19661--19668},
  year={2024},
  doi = {10.1021/acs.jpcc.4c06306},
  publisher={ACS Publications}
}

@article{HjorthLarsen_2017,
url = {https://doi.org/10.1088/1361-648X/aa680e},
year = {2017},
month = {jun},
publisher = {IOP Publishing},
volume = {29},
number = {27},
pages = {273002},
author = {Hjorth Larsen, Ask and Jørgen Mortensen, Jens and Blomqvist, Jakob and Castelli, Ivano E and Christensen, Rune and Dułak, Marcin and Friis, Jesper and Groves, Michael N and Hammer, Bjørk and Hargus, Cory and Hermes, Eric D and Jennings, Paul C and Bjerre Jensen, Peter and Kermode, James and Kitchin, John R and Leonhard Kolsbjerg, Esben and Kubal, Joseph and Kaasbjerg, Kristen and Lysgaard, Steen and Bergmann Maronsson, Jón and Maxson, Tristan and Olsen, Thomas and Pastewka, Lars and Peterson, Andrew and Rostgaard, Carsten and Schiøtz, Jakob and Schütt, Ole and Strange, Mikkel and Thygesen, Kristian S and Vegge, Tejs and Vilhelmsen, Lasse and Walter, Michael and Zeng, Zhenhua and Jacobsen, Karsten W},
title = {The atomic simulation environment—a {Python} library for working with atoms},
doi = {10.1088/1361-648X/aa680e},
journal = {Journal of Physics: Condensed Matter}
}

@article{MILLS1995305,
title = {Reversible work transition state theory: application to dissociative adsorption of hydrogen},
journal = {Surface Science},
volume = {324},
number = {2},
pages = {305--337},
year = {1995},
issn = {0039-6028},
doi = {10.1016/0039-6028(94)00731-4},
url = {https://www.sciencedirect.com/science/article/pii/0039602894007314},
author = {Gregory Mills and Hannes Jónsson and Gregory K. Schenter}
}

@article{Lindgren2024,
doi = {10.21105/joss.06264},
url = {https://doi.org/10.21105/joss.06264},
year = {2024},
publisher = {The Open Journal},
volume = {9},
number = {95},
pages = {6264},
author = {Lindgren, Eric and Rahm, Magnus and Fransson, Erik and Eriksson, Fredrik and Österbacka, Nicklas and Fan, Zheyong and Erhart, Paul},
title = {calorine: A Python package for constructing and sampling neuroevolution potential models},
journal = {Journal of Open Source Software}
}

@article{tang2009grid,
  title={A grid-based Bader analysis algorithm without lattice bias},
  author={Tang, Wei and Sanville, Eric and Henkelman, Gustavo},
  doi = {10.1088/0953-8984/21/8/084204},
  journal={Journal of Physics: Condensed Matter},
  volume={21},
  number={8},
  pages={084204},
  year={2009}
}

@article{kresse1999ultrasoft,
  title={From ultrasoft pseudopotentials to the projector augmented-wave method},
  author={Kresse, Georg and Joubert, Daniel},
  journal={Physical review b},
  doi = {10.1103/PhysRevB.59.1758},
  volume={59},
  number={3},
  pages={1758},
  year={1999},
  publisher={APS}
}

\end{document}